\documentclass[12pt]{article}
\usepackage{psfig}

\newcommand{\be}{\begin{equation}}
\newcommand{\ee}{\end{equation}}

\newcommand{\beqa}{\begin{eqnarray}}
\newcommand{\eeqa}{\end{eqnarray}}
\newcommand{\nn}{\nonumber}

\newcommand{\eqref}[1]{(\ref{#1})}

\def\CL {{\cal L}}
\def\CM {{\cal M}}

\begin{document}

\setlength{\baselineskip}{7mm}
\begin{titlepage}
\begin{flushright}
{\tt NRCPS-HE-03-08} \\

February, 2008
\end{flushright}

\vspace{1cm}
\begin{center}
{\it \Large Production of Spin-Two Gauge Bosons
}

\vspace{1cm}

{ \it{Spyros Konitopoulos} }
and
{ \it{George  Savvidy  } }

\vspace{0.5cm}

 {\it Institute of Nuclear Physics,} \\
{\it Demokritos National Research Center }\\
{\it Agia Paraskevi, GR-15310 Athens, Greece}

\end{center}

\vspace{1cm}

\begin{abstract}

\end{abstract}
We considered spin-two gauge boson production in the fermion pair
annihilation process and calculated the polarized
cross sections for each set of helicity orientations of initial and final particles.
The angular dependence of these cross sections is compared with
the similar annihilation cross sections in QED with two photons in
the final state, with two gluons in QCD and W-pair in Electroweak theory.

\end{titlepage}

\pagestyle{plain}

\begin{figure}
\centerline{\hbox{\psfig{figure=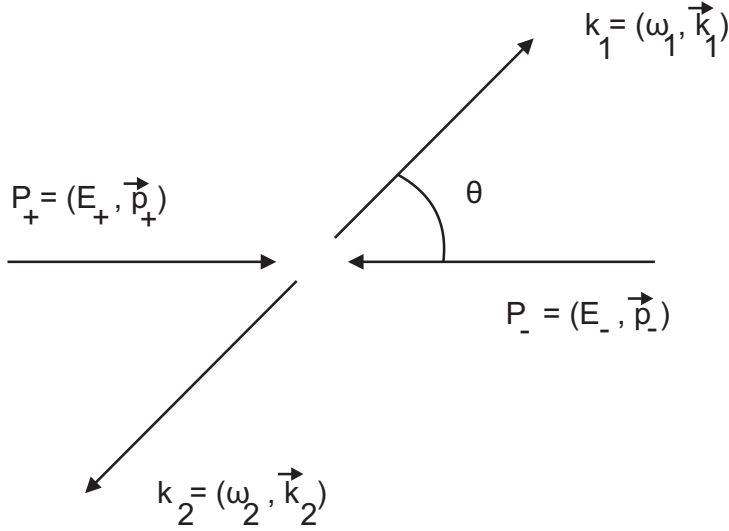,height=7cm,angle=0}}}
\caption[fig1]{The annihilation reaction $f \bar{f}  \rightarrow T T$, shown in the
center-of-mass frame. The $p_{\pm}$ are momenta of the fermions $f \bar{f}$, and $k_{1,2}$
are momenta of the tensor gauge bosons $TT$.}
\label{fig1}
\end{figure}

Our intention in this article is to calculate leading-order differential
cross section of spin-two tensor gauge boson production in the fermion pair
annihilation process $f \bar{f}  \rightarrow T T$ and to analyze the
angular dependence of the polarized cross sections
for each set of helicity orientations of initial and final particles.
The process is illustrated in   Fig.\ref{fig1}. and
receives contribution from three Feynman diagrams shown in  Fig.\ref{fig3}.
These diagrams are similar to the QED and QCD diagrams for the annihilation processes
with two photons or two gluons in the final state. The difference between these
processes is in the
actual expressions for the corresponding interaction vertices. The corresponding
vertices for spin-two tensor bosons can be found through the extension
of the gauge principle \cite{Savvidy:2005fi}.
The extended gauge principle allows to define a gauge invariant Lagrangian $\CL$ for
high-rank tensor gauge fields $A^{a}_{\mu},~A^{a}_{\mu\nu},...$
and their cubic and quartic interaction vertices
\cite{Savvidy:2005fi,Savvidy:2005zm,Savvidy:2005ki}:
\be\label{fundamentallagrangian}
{{\cal L}} =  \CL_{YM}~+~   \CL_{2}~+ ~\CL^{'} _{2}~+...
\ee

Not much is known about physical properties of similar gauge
field theories with infinite tower of fundamental fields
\cite{fierz,fierzpauli,schwinger,fronsdal,Bengtsson:1983pd,Witten:1985cc,Metsaev:2007rn}
and in the present article we shall ignore subtle aspects
of functional integral quantization procedure because we limited
ourselves to calculating only leading-order tree diagrams.
Expanding the functional integral in perturbation theory, starting with the free
Lagrangian, at $g=0$, one can see that the theory contains tensor gauge bosons
and fermions of different spins  with
cubic and quartic interaction vertices \cite{Savvidy:2005fi,Savvidy:2005zm,Savvidy:2005ki}.
Explicit form of these vertices is presented in \cite{Savvidy:2005ki}.

Below we shall present the Feynman diagrams for the given process,
the expressions for the corresponding vertices and the transition amplitude.
The transition amplitude is gauge invariant, because
if we take the physical - transverse polarization - wave function for one of the
tensor gauge bosons and unphysical - longitudinal polarization - for the second one,
the transition amplitude vanishes \cite{Konitopoulos:2008vv}.
That is unphysical - longitudinal polarization states
are not produced in the scattering process. We shall calculate
the polarized cross sections for each set of helicity orientations of the initial
and final particles
(\ref{crosssectionformulaRL-RR}), (\ref{allcrosssectionformula}) and
shall compare them with the corresponding
cross sections for photons and gluons in QED and QCD, as well as with the W-pair
production in Electroweak theory.

The annihilation process is illustrated in   Fig.\ref{fig1}.
Working in the center-of-mass frame, we make the following assignments:
$
p_- =(E_-, \vec{p}_-),~~ p_+ =(E_+, \vec{p}_+),~~ k_1 =(\omega_1, \vec{k}_1),~~
k_2 =(\omega_2, \vec{k}_2),
$
where $p_{\pm}$ are momenta of the fermions $f\bar{f} $ and $k_{1,2}$
momenta of the tensor gauge
bosons $TT$. All particles are massless $p^{2}_-=p^{2}_+  = k^{2}_1  = k^{2}_2  =0 $.
In the center-of-mass frame
the momenta satisfy the relations $\vec{p}_+  = -\vec{p}_-$, $\vec{k}_2  = -\vec{k}_1$
and $E_- = E_+ = \omega_1 = \omega_2 =E$.
The invariant variables of the process are:
\beqa
s =(p_+ + p_-)^2 = (k_1 + k_2)^2 = 2(p_+ \cdot p_-)  = 2 (k_1 \cdot k_2) \nn\\
t=(p_-  - k_1)^2 = (p_+ - k_2)^2 = -{s\over 2} (1-\cos \theta ) \nn\\
u=(p_-  - k_2)^2 = (p_+ - k_1)^2 = -{s\over 2} (1+\cos \theta ) \nn,
\eeqa
where $s= (2E)^2$ and $\theta$ is the scattering angle.

The Feynman rules for the Lagrangian (\ref{fundamentallagrangian}) can be
derived from the functional
integral over the fermion fields $\psi_{i },~\bar{\psi}_{j },~\psi^{\mu}_{i },~
\bar{\psi}^{\mu}_{j },...$ and over the gauge boson fields
$A^{a}_{\mu},~A^{a}_{\mu\nu},...$ \cite{Savvidy:2005fi,Savvidy:2005zm,Savvidy:2005ki}.
The Dirac indices are not shown, the indices of the symmetry group G are
$i,j=1,...,d(r)$, where $d(r)$ is the dimension of the representation $r$
and $a=1,...,d(G)$, where $d(G)$ is the number of generators of the group G.

\begin{figure}
\centerline{\hbox{\psfig{figure=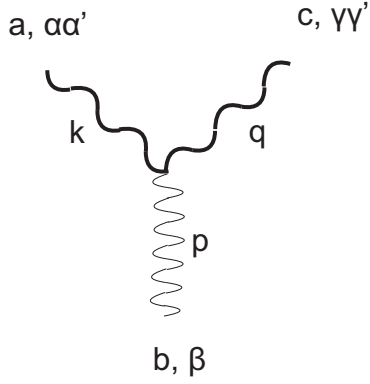,height=5cm,angle=0}}}
\caption[fig1]{The interaction vertex for vector gauge boson V and two
tensor gauge bosons T - the VTT vertex - in non-Abelian tensor gauge
field theory \cite{Savvidy:2005ki}.
Vector gauge bosons are conventionally drawn
as thin wave lines, tensor gauge bosons are thick wave lines.
The Lorentz indices $\alpha\acute{\alpha}$ and momentum $k$ belong to the
first tensor gauge boson, the $\gamma\acute{\gamma}$ and momentum $q$
belong to the second tensor gauge boson, and Lorentz index $\beta$  and
momentum $p$ belong to the vector gauge boson. }
\label{fig2}
\end{figure}

\begin{figure}
\centerline{\hbox{\psfig{figure=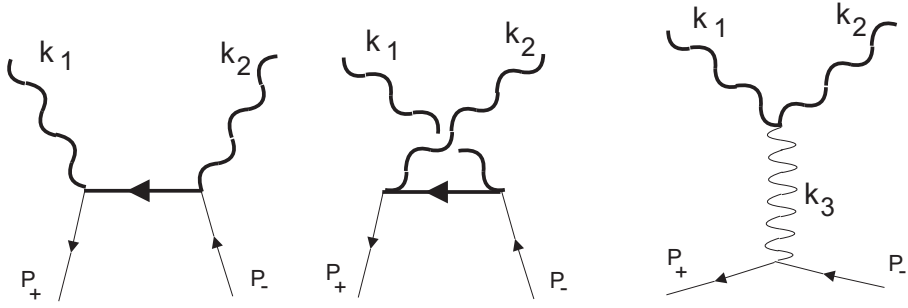,height=4cm,angle=0}}}
\caption[fig1]{Diagrams contributing to fermion-anifermion annihilation to
two tensor gauge bosons. Dirac fermions
are conventionally drawn as thin solid lines, and Rarita-Schwinger spin-vector
fermions by thick solid lines.}
\label{fig3}
\end{figure}

In the momentum space the interaction vertex of vector gauge boson V with two
tensor gauge bosons T - the VTT vertex - has the form\footnote{See formulas
(62),(65) and (66) in \cite{Savvidy:2005ki} .}
 \cite{Savvidy:2005zm,Savvidy:2005ki}
\be\label{vertexoperator}
V^{abc}_{\alpha\acute{\alpha}\beta\gamma\acute{\gamma}}(k,p,q) =
- g f^{abc} F_{\alpha\acute{\alpha}\beta\gamma\acute{\gamma}} ,
\ee
where
\beqa\label{vertexoperator1}
F_{\alpha\acute{\alpha}\beta\gamma\acute{\gamma}}(k,p,q)
&=& [\eta_{\alpha\beta} (p-k)_{\gamma}+ \eta_{\alpha\gamma} (k-q)_{\beta}
 + \eta_{\beta\gamma} (q-p)_{\alpha}] \eta_{\acute{\alpha}\acute{\gamma}}-\nn\\
-{1\over 2} \{&+&(p-k)_{\gamma}(\eta_{\alpha\acute{\gamma}}
\eta_{\acute{\alpha}\beta}+
\eta_{\alpha\acute{\alpha}} \eta_{\beta\acute{\gamma}})\nn\\
&+& (k-q)_{\beta}(\eta_{\alpha\acute{\gamma}} \eta_{\acute{\alpha}\gamma}+
\eta_{\alpha\acute{\alpha}} \eta_{\gamma\acute{\gamma}})\nn\\
&+& (q-p)_{\alpha} (\eta_{\acute{\alpha}\gamma} \eta_{\beta\acute{\gamma}}+
\eta_{\acute{\alpha}\beta} \eta_{\gamma\acute{\gamma}})\nn\\
&+&(p-k)_{\acute{\alpha}}\eta_{\alpha\beta} \eta_{\gamma\acute{\gamma}}+
(p-k)_{\acute{\gamma}} \eta_{\alpha\beta} \eta_{\acute{\alpha}\gamma}\nn\\
&+&(k-q)_{\acute{\alpha}} \eta_{\alpha\gamma} \eta_{\beta\acute{\gamma}}+
(k-q)_{\acute{\gamma}}\eta_{\alpha\gamma} \eta_{\acute{\alpha}\beta}\nn\\
&+&(q-p)_{\acute{\alpha}} \eta_{\beta\gamma} \eta_{\alpha\acute{\gamma}}+
(q-p)_{\acute{\gamma}}\eta_{\alpha\acute{\alpha}} \eta_{\beta\gamma} \}.
\eeqa
The Lorentz indices $\alpha\acute{\alpha}$ and momentum $k$ belong to the
first tensor gauge boson, the $\gamma\acute{\gamma}$ and momentum $q$
belong to the second tensor gauge boson, and Lorentz index $\beta$  and
momentum $p$ belong to the vector gauge boson. The vertex is shown in
Fig.\ref{fig2}. Vector gauge bosons are conventionally drawn
as thin wave lines, tensor gauge bosons are thick wave lines.

It is convenient to write the differential cross section in the center-of-mass frame with
tensor boson produced into the solid angle $d \Omega$ as
\be\label{crosssectionformula}
d\sigma = {1 \over 2 s} \vert M \vert^2 {1\over 32 \pi^2} d\Omega,
\ee
where the final-state density  is
$
d \Phi =   {1\over 32 \pi^2} d\Omega .
$

We shall calculate the polarized cross sections for this reaction, to lowest order
in $\alpha = g^2 / 4\pi$. The lowest-order Feynman diagrams contributing to
fermion-antifermion annihilation into a pair of tensor gauge bosons are shown in
Fig.\ref{fig3}. In order $g^2$, there are three diagrams.
Dirac fermions $\psi $
are conventionally drawn as thin solid lines, and Rarita-Schwinger spin-vector
fermions $\psi^{\mu}$ by thick solid lines.
These diagrams are similar
to the QCD diagrams for fermion-antifermion annihilation into a pair of
vector gauge bosons. The difference between these
processes is in the actual expressions for the corresponding interaction vertices
\cite{Savvidy:2005fi,Savvidy:2005zm,Savvidy:2005ki}.
The probability amplitude of the process can be written in the form
\beqa\label{matrixelement}
& \CM^{\mu\alpha\nu\beta}  e^{*}_{\mu\alpha}(k_1)e^{*}_{\nu\beta}(k_2)
= \\
&(ig)^2 \bar{v}(p_+)\{\gamma^{\mu} t^a {{1\over 4}g^{\alpha \beta } \over \not\!p_-
-\not\!k_2}t^b \gamma^{\nu}+
\gamma^{\nu}t^b  {{1\over 4}g^{\alpha \beta } \over \not\!p_-
-\not\!k_1}t^a \gamma^{\mu} +if^{abc}t^c \gamma_{\rho}{1\over k^{2}_{3}}
F^{\mu\alpha \rho \nu\beta}   \}
u(p_-) e^{*}_{\mu\alpha}(k_1)e^{*}_{\nu\beta}(k_2),\nn
\eeqa
where $u(p_-)$ is the wave function of spin $1/2$ fermion and $v(p_+)$ of antifermion,
the final tensor gauge bosons wave functions are $e^{*}_{\mu\alpha}(k_1)$
and $e^{*}_{\nu\beta}(k_2)$. The Dirac and symmetry group indices are not shown.

This amplitude is gauge invariant, that is, if we take
the physical - transverse polarization - wave function $e_T$ for one of the
tensor gauge bosons and longitudinal polarization for the second one $e_L$,
the transition amplitude vanishes: $\CM e_T e_L =0$ \cite{Konitopoulos:2008vv}.
This Ward identity expresses
the fact that the unphysical - longitudinal polarization - states
are not produced in the scattering process.

Indeed, considering the last term in (\ref{matrixelement})
and taking the polarization tensor  $e^{*}_{\nu\beta}(k_2)$ to be longitudinal
$
e^{*}_{\nu\beta}(k_2)= k_{2\nu} \xi_{\beta} + k_{2\beta} \xi_{\nu} , \nn
$
the polarization tensor  $e^{*}_{\mu\beta}(k_1)$ to be transversal and then using relations
(\ref{polarizationidentities}) for the wave function $e^{*}_{\mu\beta}(k_1)$, we shall get
\beqa\label{lasttermcontribution}
if^{abc}t^c ~\bar{v}(p_+)  \gamma^{\rho}u(p_-)~
{1\over 4}  e^{*}_{\rho\alpha}(k_1)~ \xi^{\alpha}.
\eeqa
Now let us consider the first two terms in (\ref{matrixelement}).
Taking again the polarization tensors  $e^{*}_{\nu\beta}(k_2)$ to be longitudinal
and using relations (\ref{polarizationidentities}) for the wave function $e^{*}_{\mu\beta}(k_1)$
we shall get
\beqa
&{1\over 4} \bar{v}(p_+)\{- t^a t^b \gamma^{\mu}   +
 t^b t^a  \gamma^{\mu}    \} u(p_-)
e^{*}_{\mu\alpha}(k_1) g^{\alpha \beta } \xi_{\beta} =\nn\\
& = -{1\over 4} if^{abc} t^c   \bar{v}(p_+) \gamma^{\mu}  u(p_-)
e^{*}_{\mu\alpha}(k_1)   \xi^{\alpha}  .
\eeqa
This term precisely cancels the contribution coming from the last term of the
amplitude (\ref{lasttermcontribution}).
Thus the cross term matrix element between transverse and
longitudinal polarizations vanishes: $\CM e_T e_L =0$. Our intention now is
to calculate the {\it physical matrix element $\CM e_Te_T$ for
each set of helicity orientations of initial and final particles}.

Using the explicit form of the vertex operator $F^{\mu\alpha \rho \nu\beta}$
(\ref{vertexoperator}),  (\ref{vertexoperator1}) and
the orthogonality properties of the tensor gauge boson wave functions
\beqa\label{polarizationidentities}
k^{\mu}_{1} e_{\mu \alpha}(k_1)=k^{\alpha}_{1} e_{\mu \alpha}(k_1)=
k^{\mu}_{2} e_{\mu \alpha}(k_1)=k^{\alpha}_{2} e_{\mu \alpha}(k_1)=0, \\
k^{\mu}_{2} e_{\mu \alpha}(k_2)=k^{\alpha}_{2} e_{\mu \alpha}(k_2)=
k^{\mu}_{1} e_{\mu \alpha}(k_2)=k^{\alpha}_{1} e_{\mu \alpha}(k_2)=0,\nn
\eeqa
where the last relations follow  from the fact that $\vec{k}_1  \parallel \vec{k}_2$
in the process of Fig.\ref{fig1}, we shall get
\beqa\label{polarizedtransitionamplitude}
&\CM^{\mu\alpha\nu\beta}
 e^{*}_{\mu\alpha}(k_1)e^{*}_{\nu\beta}(k_2) = (ig)^2 \bar{v}(p_+) \\
&  \{  \gamma^{\mu} t^a { {1\over 4}g^{\alpha \beta } \over \not\!p_-
-\not\!k_2}t^b \gamma^{\nu}+
\gamma^{\nu}t^b  { {1\over 4} g^{\alpha \beta } \over \not\!p_- -\not\!k_1}t^a \gamma^{\mu} +
if^{abc}t^c \gamma_{\rho}{(k_2 -k_1)^{\rho }\over k^{2}_{3}}
(g^{\mu \nu} g^{\alpha\beta} -
{1\over 2}g^{\mu\beta } g^{\nu\alpha})
 \}\nn\\
&u(p_-) ~e^{*}_{\mu\alpha}(k_1)e^{*}_{\nu\beta}(k_2). \nn
\eeqa
As the next step we shall calculate the above matrix element
in the helicity basis for initial fermions and final tensor gauge bosons.
This calculation of polarized cross sections is very similar to the one in QED \cite{feynman}.
The right- and left-handed  spinors wave functions are:
\beqa\label{helicityfermions}
u^{R}(p_-)=\sqrt{2E}
\left(\begin{array}{c}
0 \\
0 \\
0 \\
1 \\
\end{array} \right)~~~~~~~~~~~~,~~~~~~~~~~~~~~
v^{L}(p_+)=\sqrt{2E}
\left(\begin{array}{c}
0 \\
0 \\
-1 \\
0 \\
\end{array} \right)
\eeqa
and the tensor gauge boson wave functions for circular polarizations along the
$\vec{k}_1 $ direction are
\beqa\label{tensorbosonhelicities}
\epsilon_{R}^{\mu\alpha}(k_{1})={1\over 2}
 \left(\begin{array}{cccc}
0&0&0&0 \\
0&\cos^{2}\theta&i\cos\theta&-\cos\theta\sin\theta \\
0&i\cos\theta&-1&-i\sin\theta \\
0&-\cos{\theta}\sin{\theta}&-i\sin{\theta}&sin^{2}{\theta} \\
\end{array} \right),\\
\epsilon_{L}^{\mu\alpha}(k_{1})={1\over 2}
 \left(\begin{array}{cccc}
0&0&0&0 \\
0&\cos^{2}\theta&-i\cos\theta&-\cos\theta\sin\theta \\
0&-i\cos\theta&-1&i\sin\theta \\
0&-\cos{\theta}\sin{\theta}&i\sin{\theta}&sin^{2}{\theta} \\
\end{array} \right).\nn
\eeqa
It is easy to check that the wave functions (\ref{tensorbosonhelicities}) are
orthonormal
$$
\epsilon_{R}^{*\mu\alpha}(k_1)\epsilon_{L}(k_1)_{\alpha\nu }=0,~~~~
\epsilon_{R}^{*\mu\alpha}(k_1)\epsilon_{R}(k_1)_{\mu\alpha}=1~~~~,~~~~
\epsilon_{L}^{*\mu\alpha}(k_1)\epsilon_{L}(k_1)_{\mu\alpha}=1
$$
and fulfil the equations (\ref{polarizationidentities}). The helicity states
for the second gauge boson are
$
\epsilon_{R}^{\mu\nu}(k_{2})=\epsilon_{L}^{\mu\nu}(k_{1})~ ,~
\epsilon_{L}^{\mu\nu}(k_{2})=\epsilon_{R}^{\mu\nu}(k_{1}),
$
where $k_{1}^{\mu}=(E,E\sin\theta,0,E\cos\theta)$ and
$k_{2}^{\mu}=(E,-E\sin\theta,0,-E\cos\theta)$.

Now we can calculate all sixteen matrix elements between states of definite
helicities. Let us start with $\underline{f_{\tiny{R}}\bar{f}_{L}\rightarrow T_{R}T_{R}}$.
The scattering amplitude (\ref{polarizedtransitionamplitude}) for these particular helicities
$
\CM^{\mu\alpha\nu\beta}_{RL}\epsilon^{*R}_{\mu\alpha}(k_{1})\epsilon^{*R}_{\nu\beta}(k_{2})
$
contains three terms. By plugging explicit
expressions for the helicity wave functions (\ref{helicityfermions}),
(\ref{tensorbosonhelicities}) into the matrix element
(\ref{polarizedtransitionamplitude})
we can find the first term
\beqa
(ig)^2 \bar{v}^{L}(p_+)  ~  \{ \gamma^{\mu} t^a { {1\over 4}g^{\alpha \beta } \over \not\!p_-
-\not\!k_2}t^b \gamma^{\nu} \}~ u^{R}(p_-) ~e^{*R}_{\mu\alpha}(k_1)e^{*R}_{\nu\beta}(k_2)=
{(ig)^{2} \over 4}~t^{a}t^{b}~\sin\theta ,\nn
\eeqa
then the second one
\beqa
(ig)^2 \bar{v}^{L}(p_+)  ~  \{ \gamma^{\nu}t^b
{ {1\over 4} g^{\alpha \beta } \over \not\!p_- -\not\!k_1}t^a \gamma^{\mu} \}~
u^{R}(p_-) ~e^{*R}_{\mu\alpha}(k_1)e^{*R}_{\nu\beta}(k_2)=
-{(ig)^{2} \over 4}~t^{b}t^{a}~\sin\theta \nn
\eeqa
and finally the third one
\beqa
(ig)^2 \bar{v}^{L}(p_+)  \{
if^{abc}t^c  {(\not\!k_2 -\not\!k_1) \over k^{2}_{3}}
(g^{\mu \nu} g^{\alpha\beta} -
{1\over 2}g^{\mu\beta } g^{\nu\alpha})\}
u^{R}(p_-)  e^{*R}_{\mu\alpha}(k_1)e^{*R}_{\nu\beta}(k_2)=
-i{(ig)^{2} \over 2}~f^{abc}t^{c}~\sin\theta, \nn
\eeqa
so that all together they will give
\beqa\label{scatteringamplitudeRL-RR}
&\CM^{\mu\alpha\nu\beta}_{RL}\epsilon^{*R}_{\mu\alpha}(k_{1})\epsilon^{*R}_{\nu\beta}(k_{2})
={(ig)^{2} \over 4}\biggl([t^{a},t^{b}]-2if^{abc}t^{c}\biggr)~\sin{\theta}
=-i {(ig)^{2}\over 4} f^{abc} t^{c} ~\sin{\theta}.
\eeqa
To compute the cross section, we must square the matrix element
(\ref{scatteringamplitudeRL-RR}) and then average
over the symmetries of the initial fermions and sum over the
symmetries of the final tensor gauge bosons. This gives
\beqa\label{RL-RR}
\sum |\CM|^{2}_{R\bar{L}\rightarrow RR}
={g^{4} \over 16 d^{2}(r)} tr(f^{abc}f^{abd} t^{c}  t^{d} ) \sin^{2}{\theta}
={g^{4} \over 16} {C_{2}(r)C_{2}(G)\over d(r)} \sin^{2}{\theta},
\eeqa
where the invariant operator $C_2$ is defined by the equation
$ t^a t^b  = C_2 $.
Similarly, using the helicity wave functions (\ref{helicityfermions}) and
(\ref{tensorbosonhelicities}),  we can calculate the amplitude
$\underline{f_{\tiny{R}}\bar{f}_{L}\rightarrow T_{L}T_{L}}$.
This gives
\beqa\label{RL-LL}
\sum |\CM|^{2}_{R\bar{L}\rightarrow LL}
={g^{4} \over 16 d^{2}(r)} tr(f^{abc}f^{abd} t^{c}  t^{d} ) \sin^{2}{\theta}
={g^{4} \over 16} {C_{2}(r)C_{2}(G)\over d(r)} ~\sin^{2}{\theta}.
\eeqa
The amplitude  $\underline{f_{\tiny{R}}\bar{f}_{L}\rightarrow T_{R}T_{L}} $
vanishes because the common factor to all three pieces of this amplitude -
$
g^{\lambda\rho}\epsilon^{*R}_{\mu\lambda}(k_{1})\epsilon^{*L}_{\nu\rho}(k_{2})
$
- is equal to zero.
Thus only four amplitudes out of sixteen are nonzero:
\be
f_R f_{\bar{L} } \rightarrow T_R T_R,~~f_R f_{\bar{L}} \rightarrow T_LT_L,~~
 f_L f_{\bar{R}} \rightarrow T_RT_R,~~f_L f_{\bar{R}} \rightarrow T_LT_L.
\ee
From this analysis it follows that
the total spin angular momentum of the final state is one unit less than that of
the initial state, therefore a unit of spin angular momentum is converted
to the orbital angular momentum and the final state is a P-wave.

We can calculate now the leading-order polarized cross sections for the
tensor gauge boson production in the annihilation process.
Plugging matrix elements (\ref{RL-RR}) into our general
cross-section formula in the center-of-mass frame (\ref{crosssectionformula})
yields:
\beqa\label{crosssectionformulaRL-RR}
d\sigma_{f_R f_{\bar{L}} \rightarrow T_RT_R} &=&
{g^4 \over 16}~
 { C_{2}(r)~C_2(G) \over d(r)}     \sin^{2}\theta~
{1\over 2s }~{1\over 32 \pi^2} d\Omega=\nn\\
&=& {\alpha^2 \over  s   }   ~{  C_{2}(r) C_2(G) \over  64   d(r) }~
 ~  \sin^{2}\theta~ d\Omega ,
\eeqa
where
$
\alpha = {g^2 \over  4 \pi  }.
$
For the rest of the helicities we shall get
\beqa\label{allcrosssectionformula}
d\sigma_{f_R f_{\bar{L}} \rightarrow T_RT_R}=d\sigma_{f_R f_{\bar{L}} \rightarrow T_LT_L}
=d\sigma_{f_L f_{\bar{R}} \rightarrow T_RT_R}=d\sigma_{f_L f_{\bar{R}} \rightarrow T_LT_L},
\eeqa
where for the $SU(N)$ group we have ${ C_{2}(r) C_2(G) \over  64   d(r) }~
 ~=~{  (N^2 -1) \over  128 N  }$. Adding up all sixteen amplitudes and dividing
by four, to average over the initial particle spins, we recover the unpolarized cross section
\cite{Konitopoulos:2008vv}.

This cross section should be compared with the analogous annihilation cross sections
in QED and QCD. Indeed, let us compare this result with the electron-positron annihilation
into two transversal photons.
The $e^+ e^- \rightarrow \gamma \gamma$  annihilation cross section \cite{Dirac:1930}
in the high-energy limit is
\beqa\label{gammagamma}
d\sigma_{\gamma\gamma} = {\alpha^2 \over  s}{ 1+\cos^2\theta \over \sin^2\theta} d\Omega
\eeqa
except very small angles of order $m_e/E$. The cross section has a minimum
at $\theta=\pi/2$ and then increases for small
angles \cite{Duinker:1981qd}.
The quark pair annihilation cross section into two transversal gluons $q \bar{q} \rightarrow g g$
in the leading order of the strong coupling $\alpha_s$ is
\beqa\label{gluglu}
d\sigma_{gg} = {\alpha^2_s \over  s}{C_2(r)C_2(r) \over   d(r)}[ { 1+\cos^2\theta \over \sin^2\theta }
- {C_{2}(G)  \over 4 C_{2}(r)} (1+    \cos^{2}\theta)   ]d\Omega
\eeqa
and also has a minimum at $\theta=\pi/2$ and increases for small
scattering angles \cite{Abe:1993kb}. The production cross section of spin-two
gauge bosons (\ref{crosssectionformulaRL-RR}), (\ref{allcrosssectionformula}) shows dramatically
different behaviour - $\sin^2\theta$ - with its maximum at
$\theta=\pi/2$ and decrease for small angles.

It is also instructive to compare this result with the
angular dependence of the W-pair
production  in Electroweak theory. The
high energy production of {\it longitudinal } gauge bosons
is \cite{Alles:1976qv}
\beqa
d\sigma_{e^+ e^- \rightarrow W^+_0W^-_0 } =
{\alpha^2\over s }~[{1+ \sin^{4} \theta_{w}    \over 256  ~\sin^{4}\theta_{w} ~ \cos^{4}\theta_{w}  }]~
 \sin^{2}\theta~ d\Omega,
\eeqa
where $\cos\theta_{w}={m_{w}\over m_{z}}$ and it is
similar to the spin-two {\it transversal} gauge boson production
(\ref{crosssectionformulaRL-RR}).
One can only speculate that at high enough energies, may be at LHS energies and
above the threshold, we may observe the standard
spin-one gauge bosons together with new spin-two gauge bosons \cite{Savvidy:2005zm}.
To predict the threshold energy one should first construct a massive theory
and even in that case the corresponding Yukawa couplings most probably will
be unknown.

The work of (G.S.) was supported by ENRAGE (European Network on Random
Geometry), Marie Curie Research Training Network, contract MRTN-CT-2004-
005616.

\vfill
\end{document}